\documentclass[conference]{IEEEtran}
\IEEEoverridecommandlockouts

\usepackage{booktabs}

\usepackage{tabularx}
\usepackage{diagbox}

\usepackage{tikz}

\usepackage{url}

\usepackage{breakurl}
\usepackage[breaklinks]{hyperref}

\usepackage{cite}
\usepackage{amsmath,amssymb,amsfonts, amsthm}
\usepackage{algorithmic}
\usepackage{algorithm}
\usepackage{graphicx}
\usepackage{textcomp}
\usepackage{xcolor}
\usepackage{caption}
\usepackage{subcaption}
\usepackage{cleveref}

\usepackage{amssymb}
\usepackage{pifont}

\newcommand{\circo}{~\raisebox{1pt}{\tikz \draw[line width=1pt] circle(1.5pt);}~}

\def\BibTeX{{\rm B\kern-.05em{\sc i\kern-.025em b}\kern-.08em
    T\kern-.1667em\lower.7ex\hbox{E}\kern-.125emX}}

\newcounter{GeorgeNOC}
\stepcounter{GeorgeNOC}

\newcommand{\para}[1]{\vspace{1mm}\noindent\textbf{#1.}}
\newcommand{\emphPara}[1]{\vspace{1mm}\noindent\textit{-- #1.}}

\newtheoremstyle{bfnote}%
{}{}%
{}{}%
{\bfseries}{.}%
{ }%
{\thmname{#1}\thmnumber{ #2}\thmnote{ (#3)}}

\theoremstyle{bfnote}
\newtheorem{definition}{Definition}

\definecolor{asparagus}{rgb}{0.53, 0.66, 0.42}

\newcounter{AsteriosNOC}
\stepcounter{AsteriosNOC}

\newcounter{KyriakosNOC}
\stepcounter{KyriakosNOC}

\newcounter{MariosNOC}
\stepcounter{MariosNOC}

\newcounter{ParisNOC}
\stepcounter{ParisNOC}

\makeatletter
\newcommand{\linebreakand}{%
  \end{@IEEEauthorhalign}
  \hfill\mbox{}\par
  \mbox{}\hfill\begin{@IEEEauthorhalign}
}
\makeatother

\begin{document}

\pagestyle{plain}

\title{CheckMate: Evaluating Checkpointing Protocols \\ for Streaming Dataflows}

\author{

\IEEEauthorblockN{George Siachamis}
\IEEEauthorblockA{
\textit{Delft University of Technology}\\
g.siachamis@tudelft.nl}
\and
\IEEEauthorblockN{Kyriakos Psarakis}
\IEEEauthorblockA{
\textit{Delft University of Technology}\\
k.psarakis@tudelft.nl}
\and
\IEEEauthorblockN{Marios Fragkoulis}
\IEEEauthorblockA{\textit{Delft University of Technology} \\ m.fragkoulis@tudelft.nl}
\linebreakand
\IEEEauthorblockN{Arie van Deursen}
\IEEEauthorblockA{
\textit{Delft University of Technology}\\
arie.vandeursen@tudelft.nl}
\and
\IEEEauthorblockN{Paris Carbone}
\IEEEauthorblockA{\textit{KTH Royal Institute of Technology} \\
parisc@kth.se}
\and
\IEEEauthorblockN{Asterios Katsifodimos}
\IEEEauthorblockA{
\textit{Delft University of Technology}\\
a.katsifodimos@tudelft.nl}

}

\maketitle

\begin{abstract}
Stream processing in the last decade has seen broad adoption in both commercial and research settings. One key element for this success is the ability of modern stream processors to handle failures while ensuring exactly-once processing guarantees. At the moment of writing, virtually all stream processors that guarantee exactly-once processing implement a variant of Apache Flink's coordinated checkpoints  -- an extension of the original Chandy-Lamport checkpoints from 1985. However, the reasons behind this prevalence of the coordinated approach remain anecdotal, as reported by practitioners of the stream processing community. At the same time, common checkpointing approaches, such as the uncoordinated and the communication-induced ones, remain largely unexplored.

This paper is the first to address this gap by $i)$ shedding light on why practitioners have favored the coordinated approach and $ii)$ by investigating whether there are viable alternatives. To this end, we implement three checkpointing approaches that we surveyed and adapted for the distinct needs of streaming dataflows. Our analysis shows that the coordinated approach outperforms the uncoordinated and communication-induced protocols under uniformly distributed workloads. To our surprise, however, the uncoordinated approach is not only competitive to the coordinated one in uniformly distributed workloads, but it also outperforms the coordinated approach in skewed workloads. We conclude that rather than blindly employing coordinated checkpointing, research should focus on optimizing the very promising uncoordinated approach, as it can address issues with skew and support prevalent cyclic queries. We believe that our findings can trigger further research into checkpointing mechanisms.
\end{abstract}

\section{Introduction}
Streaming queries constitute a crucial component of cloud applications, such as online advertising, fraud detection, real-time analytics, and Internet of Things (IoT) use cases. Streaming queries are commonly executed within multi-tenant distributed environments, subject to varying service level agreements (SLAs) regarding fault-tolerance, processing guarantees (e.g., at-least/exactly-once processing), and uptime. 

The first generations of streaming engines delegated the responsibility of correctness mechanisms to the application programmers~\cite{abadi2005borealis, arasu2003stream, sirish2003telegraphcq}. With the advent of cloud computing, modern streaming engines, such as Apache Flink~\cite{carbone2015flink}, Google Millwheel~\cite{akidau2013millwheel}, SEEP~\cite{castro2013seep}, IBM Streams~\cite{silva2016ibm}, Hazelcast Jet~\cite{hazelcast2021jet}, and Microsoft Trill~\cite{badrish2015trill} have adopted more advanced fault tolerance mechanisms, that achieve exactly-once processing guarantees~\cite{carbone2017flinkstate,silvestre2021clonos}, without the need for programmers to change the business logic to cater for failures.

At the moment of writing, there is consensus in the use of the classic coordinated checkpointing protocol~\cite{chandyLamport1985coordinated} and its variants for rollback recovery across production-grade stream processing engines, following its initial undertaking in Apache Flink~\cite{carbone2017flinkstate}. Coordinated checkpointing protocols leverage special messages, known as markers, to capture a consistent checkpoint of the distributed global state in a coordinated fashion. Once a failure occurs, a streaming pipeline can recover by rolling all operators back to their latest checkpoint and resuming processing from an offset of the streaming input.

Despite its wide adoption, the coordinated approach has been criticized for two main drawbacks. The first is that, in large deployments, the coordination can block operators with a large number of inputs (e.g., joins or aggregates) during the marker alignment phase \cite{largeShufflesDocumentation,largeShufflesDocumentationAlibaba}. The second issue is that in case of backpressure \cite{backpressureCheckpointsBlog,backpressureCheckpointsBlogAmazon}, the markers cannot travel through the dataflow graph, and the checkpointing mechanism stalls, eventually halting the processing of new messages.

At the same time, multiple approaches have been proposed in the past, stemming from the original uncoordinated \cite{bhargava1987independent,wang1995rollbackAlgorithm} and communication-induced \cite{alvisi1999cicexp,briatico1984BCS,hellary2000HMNR} checkpoints (CIC). Uncoordinated protocols allow processes to take checkpoints independently, without coordination via markers, but $i)$ they require storing logs of in-flight messages, $ii)$ they need to execute a recovery-line algorithm before recovery, and $ii)$~the number of messages that need to be replayed upon recovery can be substantially large (depending on the recovery line found). To alleviate these issues, the communication-induced family of protocols can limit the rollback propagation during recovery by breaking the patterns that lead to invalid checkpoints with forced checkpoints during normal execution.  

Despite this convergence of the stream processing engines to the coordinated checkpointing protocol, no substantial experimental evidence currently supports this system design decision against other options (e.g., uncoordinated and communication-induced checkpoints). This lack of experimental evidence can lead future streaming engines to adopt the predominant coordinated protocol along with its drawbacks, while alternative options that could behave better are ignored. Therefore, further investigation is crucial to facilitate both research and practice toward classifying checkpointing protocols and reasoning about the protocol choices that meet the needs of different workloads.

In addressing these gaps, this work is the first to revisit checkpointing for stream processing from its first principles. First, we present and analyze the theoretical cost of existing approaches. We then experimentally evaluate the three prominent checkpointing protocol families by implementing them in a testbed system built for the needs of this evaluation. We push the protocols to their limits on diverse workloads, resulting in various topologies and processing needs, including a cyclic query. Finally, we measure the performance and the impact of the protocols both on failure-free execution as well as under failure in both uniform and skewed workloads. 

\noindent In summary, this paper makes the following contributions:

\begin{itemize}
    \item A comprehensive survey of three families of checkpointing approaches and the conditions under which they can guarantee exactly-once processing.
    \item A theoretical account of the advantages and drawbacks of those three checkpointing approaches in streaming dataflows.
    \item An open-source streaming dataflow testbed system that enables accurate and isolated comparison of different checkpointing protocols.
    \item The first experimental evaluation of three checkpointing approaches on different workloads using NexMark queries~\cite{tucker2008nexmark} and a custom query that causes cycles in the dataflow graph.  
    \item The first experimental evidence showing that:
        \begin{itemize}
            \item Under \textit{uniformly} distributed workloads, the coordinated approach  outperforms all other approaches;
            \item Under \textit{skewed} workloads, the uncoordinated approach outperforms the coordinated one despite its expensive in-flight message logging;
            \item The uncoordinated approach in practice does not suffer from the (theoretical) domino effect~\cite{elnozahy2002surveyRecovery} in any of our experiments.
            \item The communication-induced approach is not competitive in any scenario due to its large message overhead that it requires to avoid the (improbable, in our experiments) domino effect.
        \end{itemize}
\end{itemize}

The rest of the paper is structured as follows. In \Cref{sec:preliminaries}, we summarise all the necessary background knowledge required to understand checkpointing. Then we discuss in detail the benchmarked protocols (\Cref{sec:checkpointing-protocols}) and the system used for the benchmarking (\Cref{sec:system}). In \Cref{sec:experiments}, we describe the experimental setup and present and comment on our results. In \Cref{sec:related-work}, we discuss related existing works. \Cref{sec:conclusions} concludes this paper.

\vspace{1mm}
\noindent The code of CheckMate can be found online:\\ \url{https://github.com/delftdata/checkmate}

\section{Preliminaries}
\label{sec:preliminaries}

In what follows, we discuss all the necessary concepts to understand better and evaluate the checkpointing protocols, particularly processing semantics and consistency in the face of failures.

\begin{figure}[t!]
    \centering
    \includegraphics[width=0.75\linewidth]{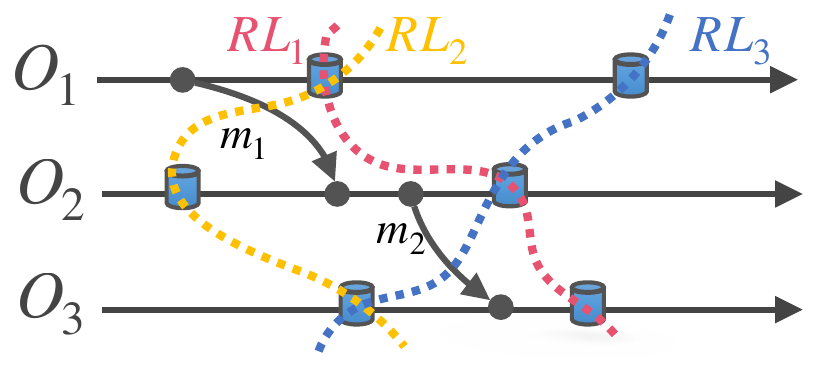}
    \caption{Examples of valid recovery lines when in-flight messages are included in the global state.}
    \label{fig:rec-lines}
    \vspace{-3mm}
\end{figure}

\subsection{Processing Semantics}
Different applications have different processing needs. Stream processing engines and their fault tolerance mechanisms provide specific processing semantics to accommodate these needs even when a failure occurs. A recent survey~\cite{fragkoulis2023survey} identifies three predominant semantics with regard to processing: \emph{at-most-once}, \emph{at-least-once}, and \emph{exactly-once}.

For data analytics, monitoring, or other applications that can tolerate incomplete data, a stream processing engine that provides \emph{at-most-once semantics} is sufficient. We define \emph{at-most-once} semantics as follows:

\begin{definition}[At-most-once]
A stream processing engine provides \emph{at-most-once} processing semantics when it ensures that each streaming operator will process each ingested record once or not at all.
\end{definition}

At-most-once semantics are the weakest guarantees a stream processing engine can provide. This processing guarantee has been termed \textit{gap recovery} in the past~\cite{Hwang05high}. In case of a failure, in-flight records can be lost and never be processed by downstream operators. 

To accommodate applications that are intolerant of losing messages, streaming engines support \emph{at-least-once} semantics.

\begin{definition}[At-least-once]
\emph{At-least-once} processing semantics are provided when each ingested message is processed one or more times by each streaming operator. 
\end{definition}

By providing at-least-once semantics, a streaming engine can avoid data loss, but at the same time, it is amenable to accounting for the same message more than once. For sensitive applications, such as bank transaction handling or aggregations, duplicate processing can cause serious anomalies. In such cases, \emph{exactly-once} semantics are necessary.

\begin{definition}[Exactly-once]
\emph{Exactly-once} semantics guarantee that each ingested message is processed exactly once in each operator, i.e., any state changes that occur from processing a message are reflected exactly-once on the checkpointed state. 
\end{definition}

Exactly-once semantics define strict guarantees, and they can ensure that processing under failures is identical to failure-free processing. Note that there is a distinction \cite{fragkoulis2023survey} between exactly-once \textit{processing} and exactly-once \textit{output} \cite{damani1996recover}. In exactly-once processing, an external system consuming the output can still observe duplicates. For instance, in case of fault recovery, the streaming system will resume processing after the latest checkpoint and produce some output that it had already produced (but not yet checkpointed the corresponding state) prior to the failure.

In the rest of the paper, we only consider \emph{exactly-once} processing guarantees. 

\subsection{Consistency of Global State}

Data stream execution is data-driven, where processing is orchestrated by messages being sent and received between tasks, triggering local computation. Without loss of generality, a distributed stream execution consists solely of \texttt{send} and \texttt{receive} operations corresponding to each message.

Modern distributed stream processing engines refer to the \emph{global state} as the collection of the states of all operators of a streaming pipeline. We refer to an operation (\texttt{send} or \texttt{receive}) being part of the global state if it occurred before the respective state acquisition. Furthermore, the state of the communication channels can also be included in the global state. These messages are also known as \emph{in-flight messages} or \emph{channel state}. The \emph{consistency}~\cite{chandyLamport1985coordinated} of the global state is of major importance here. In order to define what a consistent state entails, we first define the concept of orphan messages:

\begin{definition}[Orphan message]
    Given a global state checkpoint G, an \textit{orphan message} has been received prior to the receiver's local checkpoint S in G, but it was not sent prior to the sender's checkpoint S' in G.
\end{definition}

The global state of a streaming pipeline becomes inconsistent in the presence of a dropped or an orphan message \cite{strom1985orphan, cao1998coordinatedCheck, elnozahy2002surveyRecovery}. Following the seminal processing model of Chandy-Lamport~\cite{chandyLamport1985coordinated}, we define consistent global state as follows:

\begin{definition}[Consistent global state]
    The global state $G$ of a streaming pipeline is \emph{consistent} if for each message $m$ :
        \begin{itemize}
            \item \textbf{No Orphans:} if \texttt{receive(m)} happened before the checkpoint   acquisition, the corresponding \texttt{send(m)} operation should also happen before the checkpoint.
            \item \textbf{No Dropping:} if \texttt{send(m)} happened before the checkpoint acquisition then either \texttt{receive(m)} happens before the checkpoint or $m$ is added in the checkpoint as an \emph{in-flight} message.
        \end{itemize} 

\end{definition}

In principle, consistency is straightforward to maintain and reason about under the normal operation of a streaming system. In the face of failures, however, a streaming system ought to roll back to a previously consistent global state in order to resume its operation and regain consistency. At that point, the recovery mechanism attempts to recover such a global state from the collection of existing operator checkpoints.

\begin{table*}[ht!]
\begin{tabularx}{\textwidth}{l|ccc|ccccc}
                               & \textbf{\begin{tabular}[c]{@{}c@{}}Blocking \\ (markers)\end{tabular}} & \textbf{\begin{tabular}[c]{@{}c@{}}In-flight \\ Logging\end{tabular}} &\textbf{\begin{tabular}[c]{@{}c@{}}Deduplication \\ Required \end{tabular}}  & \textbf{\begin{tabular}[c]{@{}c@{}}Message \\ Overhead\end{tabular}} & \textbf{\begin{tabular}[c]{@{}c@{}}Independent \\ Checkpoints\end{tabular}}& \textbf{\begin{tabular}[c]{@{}c@{}}Straggler \\ Stalls\end{tabular}}& \textbf{\begin{tabular}[c]{@{}c@{}}Unused \\ Checkpoints\end{tabular}} & \textbf{\begin{tabular}[c]{@{}c@{}}Forced \\ Checkpoints\end{tabular}}                                        \\ \toprule
\textbf{Coordinated}           & \circo  & ~--      & ~--     & ~--    & ~--     & \circo & ~-- & ~--\\ \midrule
\textbf{Uncoordinated}         & ~--     & \circo   & \circo  & ~-- & \circo  & ~-- & \circo & ~-- \\ \midrule
\textbf{Communication-induced} & ~--     & \circo   & \circo  & \circo & \circo  & ~-- & \circo & \circo \\\bottomrule                                                         
\end{tabularx}
\caption{Summary of the features of the checkpointing protocols explored in \Cref{sec:checkpointing-protocols}}
\label{tab:overview}
\vspace{-2mm}
\end{table*}

\para{Recovery line}
A \textit{recovery line} consists of a collection of operator checkpoints that can be used to recover the global state (\Cref{fig:rec-lines}). Since not all candidate recovery lines lead to a consistent state, the recovery mechanism must find the most recent recovery line corresponding to a consistent state. Checkpoints that cannot belong to a consistent recovery line are considered \emph{invalid}.

\begin{figure}[t]
    \centering
    \begin{subfigure}[h]{\linewidth}
    \centering
        \includegraphics[width=0.75\linewidth]{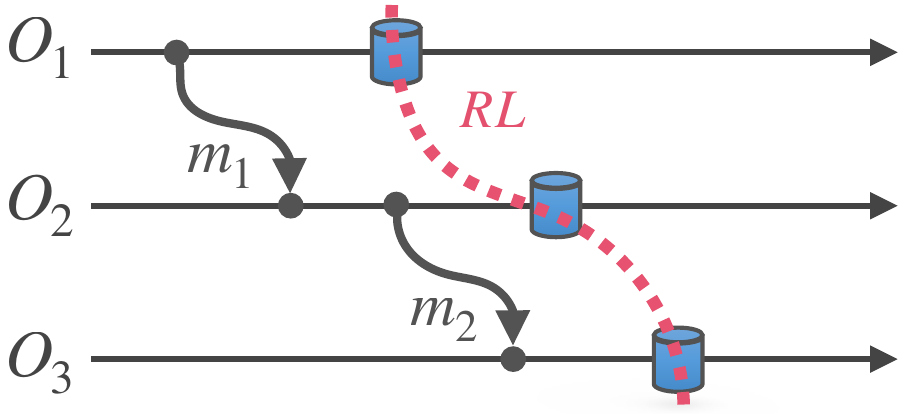}
        \caption{Consistent state after recovery}
        \label{fig:state-consistency-a}
    \end{subfigure}
    
    \begin{subfigure}{\linewidth}
    \centering
        \includegraphics[width=0.75\linewidth]{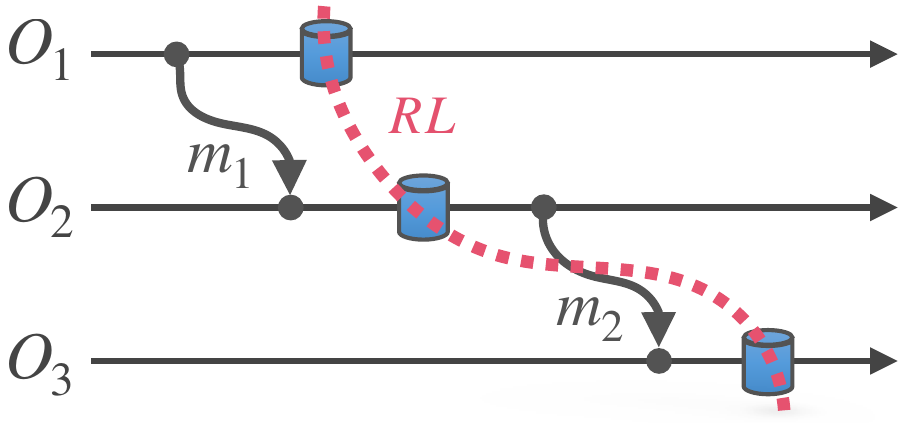}
        \caption{Inconsistent state after recovery}
        \label{fig:state-consistency-b}
    \end{subfigure}
    
    \begin{subfigure}{\linewidth}
    \centering
        \includegraphics[width=0.75\linewidth]{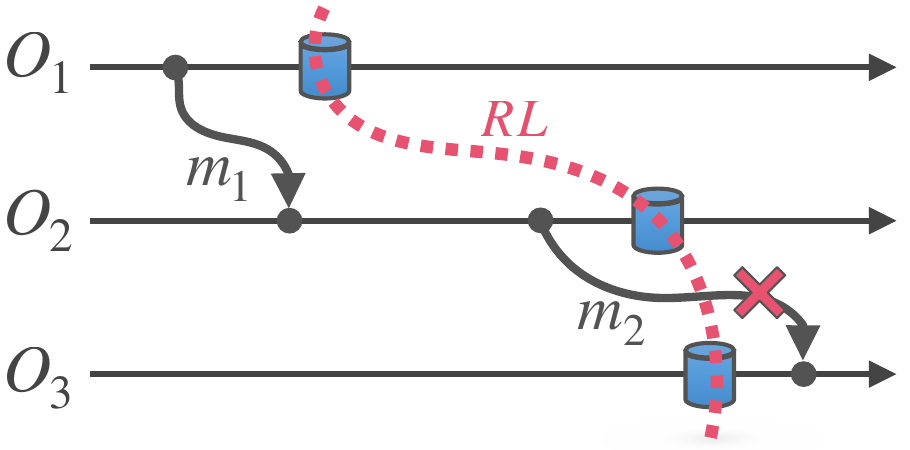}
        \caption{Inconsistent state without capturing in-flight messages}
        \label{fig:state-consistency-c}
    \end{subfigure}

    \begin{subfigure}{\linewidth}
    \centering
        \includegraphics[width=0.75\linewidth]{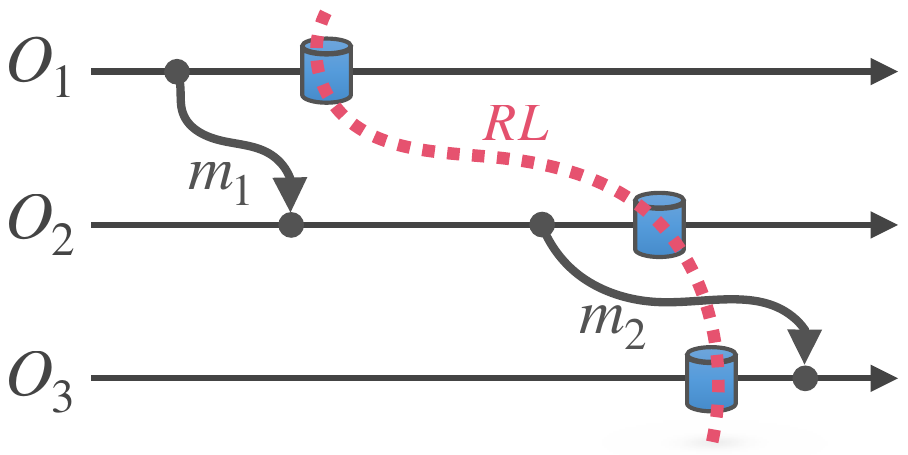}
        \caption{Consistent state by capturing in-flight messages}
        \label{fig:state-consistency-d}
    \end{subfigure}
    \caption{Cases of inconsistent and consistent state after recovery for stateful operators $O_1, O_2$ and $O_3$.}
    \label{fig:state-consistency}
    \vspace{-4mm}
\end{figure}

In \Cref{fig:state-consistency}, we provide example cases that illustrate when a recovery line and its corresponding global state are consistent. \Cref{fig:state-consistency-a} showcases a consistent global state since all messages are sent and received before the checkpoints that compose the recovery line. In \Cref{fig:state-consistency-b}, message $m_2$ is an orphan message since its side-effects are reflected in the checkpoint of $O_3$ but not in the checkpoint of the sender operator $O_2$. Therefore, the global state corresponding to this recovery line is inconsistent, and the recovery line is unsuitable for recovering from a failure. 

If in-flight messages (i.e., channel state) are not captured, then a different type of global state inconsistency appears. In \Cref{fig:state-consistency-c}, operation \texttt{send($m_2$)} occurs before $O_2$ acquires its checkpoint, whereas, \texttt{receive($m_2$)} occurs after $O_3$ takes its checkpoint. Using this recovery line without a captured channel state will result in never processing $m_2$ at operator $O_3$ and, therefore, dropping messages. In this case, to achieve a consistent global state, capturing the channel state and replaying in-flight messages is necessary (\Cref{fig:state-consistency-d}). To ensure exactly-once semantics when in-flight messages are replayed, some form of message deduplication must be employed.

\section{Checkpointing Protocols}
\label{sec:checkpointing-protocols}

In what follows, we describe the three main checkpointing protocols and discuss their core ideas and some possible drawbacks. In \cref{tab:overview}, we summarise the necessary mechanisms employed for each protocol to ensure exactly-once processing and the main side effects and features of each protocol.

\begin{figure*}[t]
    \centering
    \includegraphics[width=\textwidth]{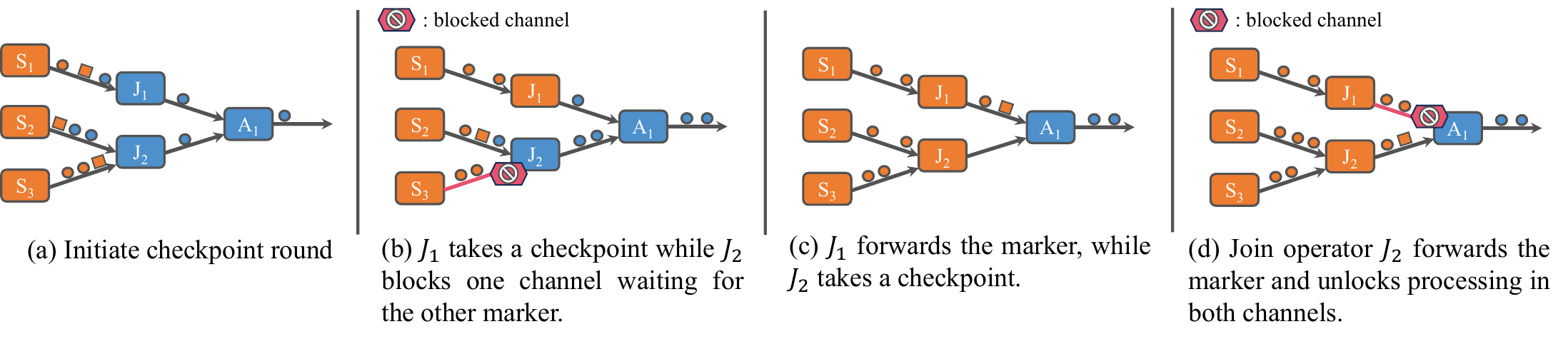}
    \caption{Example execution of the coordinated aligned checkpointing protocol. Messages are represented as circles,
and markers are squares. Different colors denote different coordinated rounds.}
    \label{fig:coordinated}
    \vspace{-1mm}
\end{figure*}

\subsection{Coordinated Aligned Checkpointing (COOR)}

To the best of our knowledge, virtually every stream processing engine in production that guarantees exactly-once processing, implements a variation of the coordinated checkpointing protocol \cite{carbone2017flinkstate, hazelcast2021jet, badrish2015trill, silva2016ibm}. Typically, in stream processing engines that implement a coordinated checkpointing protocol, the operators will block processing to allow the alignment of a checkpoint across the system. The checkpoint can be used to create a recovery line in case of a failure. The most adopted version of such a protocol is the Chandy-Lamport marker-based algorithm~\cite{chandyLamport1985coordinated} and its adaptation for acyclic dataflow graphs~\cite{carbone2017flinkstate}. In what follows, we describe the core ideas of the protocol, and we illustrate its core functionality with an example.

At its core, the coordinated aligned checkpointing protocol works as follows:
\begin{itemize}
    \item A checkpoint round initiates at source operators by taking a checkpoint. After taking its checkpoint, each source operator forwards a marker to all its outgoing channels and continues processing.
    \item When an operator (excluding source operators) receives a marker from an incoming channel, it blocks that channel and buffers the channel's traffic.
    \item When an operator receives a marker from all its incoming channels, it takes a checkpoint, unblocks processing in all incoming channels, and forwards a marker to all its outgoing channels.
    \item When the markers reach the end of the pipeline, and the checkpoints are stored in durable storage, the coordinated checkpoint round finishes.
\end{itemize}

By blocking processing until all markers are received from the upstream operators, we achieve the alignment of the checkpoints. This alignment guarantees exactly-once processing without the need to capture in-flight messages and the channel state, as it creates a frontier of processed messages through the use of markers.

\Cref{fig:coordinated} illustrates an example protocol execution. The execution graph presented consists of only the first couple operators of the pipeline. Operators $S_{\{1-3\}}$ are parallel source operators, operators $J_1$ and $J_2$ are parallel stateful join operators, and operator $A_1$ is a stateful aggregation operator. A coordinated checkpoint round is initiated at the source operators by taking a checkpoint. When a parallel source operator finishes with its own checkpoint, it sends a checkpointing marker to all its outgoing channels (\cref{fig:coordinated}(a)) and continues processing. In \cref{fig:coordinated}(b), operator $J_1$ has received the marker from its sole incoming channel and takes a checkpoint. On the other hand, operator $J_2$ has received a marker from source operator $S_3$ and blocks processing in that channel while it waits for the marker from $S_2$. $J_2$ takes a checkpoint when it has received all markers, while $J_1$, after taking the checkpoint, forwards a marker to its downstream operator and unblocks processing in all the incoming channels (\cref{fig:coordinated}(c)). Finally, $J_2$ also forwards a marker and continues processing after taking a checkpoint (\cref{fig:coordinated}(d)). The markers will then be received by $A_1$, and the checkpointing process will continue in the same way until it reaches the end of the pipeline.

\para{Strengths} Compared to the in-flight message logging required in uncoordinated approaches (\Cref{sec:uncoordinated}), the markers used by the coordinated protocol are lightweight and are not affected by the message size. Additionally, since aligned checkpoints compose a consistent global state, an algorithm that identifies the recovery line is not required.

\para{Drawbacks} One important downside of marker circulation surfaces in cases of stragglers, e.g., due to skewed workloads and/or backpressure. For example, if most of the load falls on a single operator, its downstream operators would have to block the other channels, wait for the straggler to finish processing, and then forward a checkpoint marker. Additionally, in case of shuffling, the protocol needs to transfer as many markers as the parallel instances of the receiving operators (one to each parallel instance). In essence, coordinated checkpoints could take a substantial amount of time in complex topologies due to the markers having to pass through the entire dataflow graph to be completed.

Another drawback of the coordinated protocol is that it does not support cyclic streaming workloads out of the box. Cycles are an integral aspect of iterative computations such as fixpoint calculations, which are common in graph queries~\cite{mcsherry2013naiad}. Accounting for cycles in the coordinated checkpointing protocol entails a) special handling of markers in order to avoid deadlocks owed to the blocking of the cyclic input channel by a marker and b) additional progress tracking mechanisms. 

\vspace{-1mm}
\subsection{Uncoordinated Checkpointing (UNC)}
\label{sec:uncoordinated}

The \emph{uncoordinated checkpointing} (UNC)~\cite{wang1995rollbackAlgorithm} protocol allows each operator to decide individually when to take a checkpoint. In contrast to the coordinated approach, there are no markers since there is no need for coordination, and the protocol can only provide at-most-once processing semantics since the checkpoints only contain the operator state. Thus, capturing the channel state between operators is necessary to provide stronger guarantees. To do so, log-based recovery and upstream backup~\cite{balazinka2017upstream,cherniak2003upstream} need to be implemented. Pairing uncoordinated checkpointing with a log for keeping track of the channel state allows the replay of messages after recovery, achieving at-least-once semantics. For the protocol to achieve exactly-once semantics, message \textit{deduplication} must be employed when replaying messages from the message log.

\begin{algorithm}[h!]
\caption{Rollback propagation algorithm \cite{wang1995rollbackAlgorithm}}\label{alg:rollback}
\begin{algorithmic}[1]

\REQUIRE all available checkpoints $CP$ ordered by freshness for each operator, a checkpoints graph
\ENSURE a consistent recovery line
\STATE include in $root\_set$ the latest $CP$ of each operator;
\STATE mark all $CPs$ in the $root\_set$ that are strictly reachable from any other $CP$ in the $root\_set$;
\WHILE{$\exists CP.marked \in root\_set$}
\STATE $\forall CP.marked \in root\_set$ replace by the next unmarked $CP$ from the same operator;
\STATE mark all $CPs$ in the $root\_set$ that are strictly reachable from any other $CP$ in the $root\_set$;
\ENDWHILE
\RETURN $root\_set$
\end{algorithmic}
\end{algorithm}

\para{Finding Recovery Lines} The freedom of taking checkpoints independently per operator comes with a cost when recovering after a failure. Since the checkpoints are not coordinated, we cannot simply use the most recent operator checkpoints as a recovery line, as it might not correspond to a consistent global state. Therefore, we need to employ a recovery line algorithm to find a suitable recovery line, i.e., one that provides a consistent global state and has the minimum rollback distance. The algorithm for finding such a recovery line is the \emph{rollback propagation algorithm}~\cite{wang1995rollbackAlgorithm}, which requires a checkpoint dependency graph. There are two approaches to creating such a graph, the \emph{rollback dependency graph}~\cite{bhargava1988rollbackgraph} and the \emph{checkpoint graph}~\cite{wang1995rollbackAlgorithm}. Both of these approaches result in the same recovery line, and in this work, we opt for the \emph{checkpoint-graph}~\cite{wang1995rollbackAlgorithm} since it is more intuitive. 

The checkpoint graph has checkpoints as nodes and directed edges between two checkpoints $c_{i,x}$ and $c_{j,y}$ if:
\begin{itemize}
    \item $i \neq j$, i.e., the checkpoints belong to different operators, and there is at least one orphan message that was sent from operator $i$ after checkpoint $c_{i,x}$ was captured and was processed from operator $j$ before checkpoint $c_{j,y}$ was taken.
    \item $i = j$ and $y=x+1$, i.e., $c_{i,x}$ and $c_{j,y}$ are consecutive checkpoints of the same operator.
\end{itemize}

\begin{figure*}[t]
    \centering
    \includegraphics[width=\linewidth]{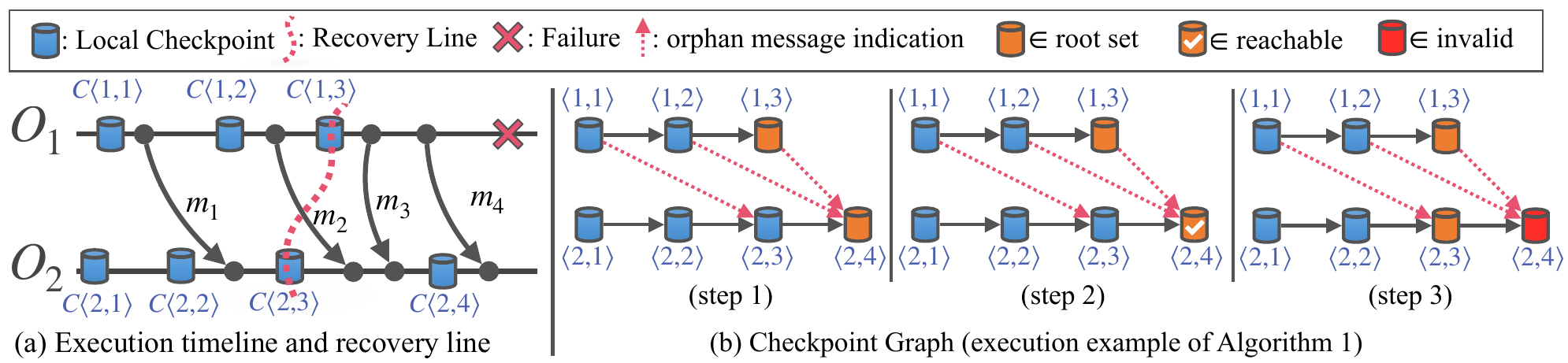}
    \caption{Example overview of Rollback propagation algorithm on a given execution timeline}
    \label{fig:checkpoint_graph}
    \vspace{-3mm}
\end{figure*}

In \Cref{fig:checkpoint_graph}, we provide an example of a checkpoint graph and showcase step by step how the rollback propagation algorithm uses the checkpoint graph to find a suitable recovery line. To create the checkpoint graph, we include the IDs from channel state logs for the last received and last sent messages alongside the checkpoints. We can identify orphan messages using these IDs and add directed edges in the checkpoint graph~(\Cref{fig:checkpoint_graph}(a)). The rollback propagation algorithm uses this graph to find the recovery line. First, the algorithm will include the last checkpoints of all operators in a set called the root set (\Cref{fig:checkpoint_graph}(b) - step 1). The next step is to identify the nodes in the root set that are strictly reachable from other nodes in the root set and mark them (\Cref{fig:checkpoint_graph}(b) - step 2). Then, each marked checkpoint in the root set is replaced by the next most fresh checkpoint for the same operator, and the newly added checkpoints are checked and marked if applicable (\Cref{fig:checkpoint_graph}(b) - step 3). When the algorithm reaches a root set that does not include any marked checkpoint, it returns this root set as the desired recovery line.

\para{Strengths} The primary strength of any coordination-free protocol is that it does not block waiting for markers from a coordinator node or its upstream operators, leading to lower latency in the event of a skewed workload. Another benefit yet to be explored by literature is the configurability of such an approach. For instance, the stateless, non-source operators in the uncoordinated approach do not need to participate in the checkpointing pipeline, which is not the case in the coordinated approach because they still would have to propagate the markers. Furthermore, different operators can have different checkpoint intervals, making them adaptive to the current system's needs (e.g., a windowed aggregation operator can checkpoint right after the aggregate is calculated in order to avoid storing the large window's contents).

\begin{figure}[t!]
    \centering
    \includegraphics[width=0.8\linewidth]{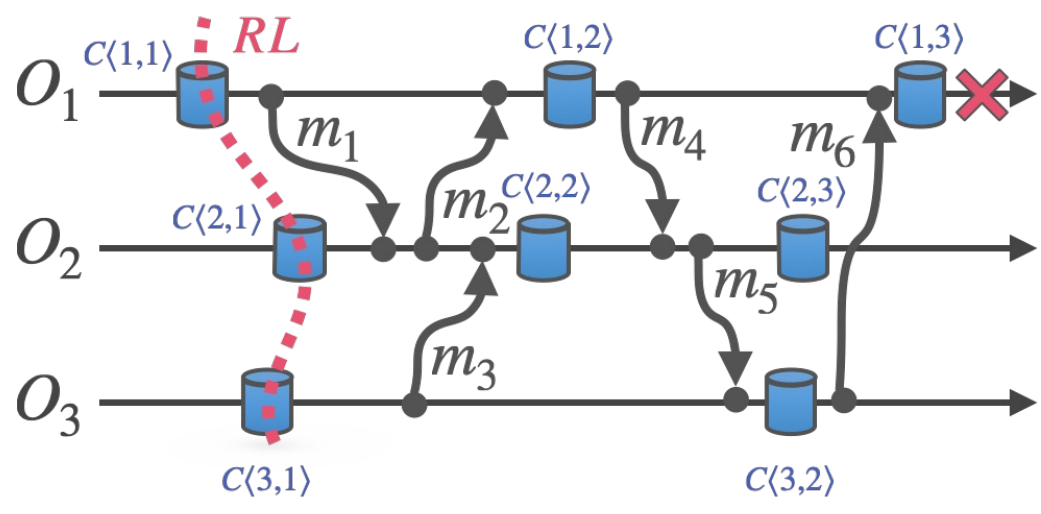}
    \caption{Domino effect of invalid checkpoints on a cyclic query.}
    \label{fig:domino}
    \vspace{-3mm}
\end{figure}

\para{Drawbacks} To provide exactly-once semantics, message logging is required. However, message logging is costly and can considerably impact the system's performance. Moreover, since checkpoints are not aligned, some captured checkpoints may be rendered invalid when looking for the appropriate recovery line (an invalid checkpoint cannot take part in any recovery line). As seen in \Cref{fig:domino}, this problem could be aggravated when dealing with cyclic queries, leading to a phenomenon known in the literature as the \emph{unbounded domino effect}~\cite{elnozahy2002surveyRecovery}, where during recovery, one checkpoint after the other is rendered invalid leading to a considerable rollback distance or even starting from scratch. In \Cref{fig:domino}, the first option would be a recovery line consisting of the checkpoints  $C_{<1,3>}$, $C_{<2,3>}$, and $C_{<3,2>}$; however, this is invalid due to the orphan message $m_6$. The next option is the recovery line consisting of $C_{<1,2>}$, $C_{<2,3>}$, and $C_{<3,2>}$ with again $m_4$ making this invalid. $m_5$ makes the $C_{<1,2>}$, $C_{<2,2>}$, and $C_{<3,2>}$ invalid. The domino effect continues with the rest of $m_{3,2,1}$ leading to $C_{<1,1>}$, $C_{<2,1>}$, and $C_{<3,1>}$ being the only available recovery line option.

\subsection{Communication-induced Checkpointing (CIC)}

The \emph{communication-induced checkpointing} (CIC) protocol is built on top of UNC and provides a loose coordination of the checkpoints in order to tackle the problem of the \emph{unbounded domino effect}. This loose coordination happens through encapsulating information related to the protocol in the messages containing records across the pipeline. This protocol recognizes two different types of checkpoints: a) \textit{local checkpoints} (equivalent to uncoordinated checkpoints), and b) \textit{forced checkpoints}, which are inserted by the protocol to prevent the domino effect. 

Communication-induced protocols are tightly connected to Z-paths and Z-cycles~\cite{elnozahy2002surveyRecovery} based on the fact that a given checkpoint is invalid if and only if it is part of a Z-cycle. A CIC protocol tries to detect Z-cycles and break them by forcing checkpoints before processing messages that will lead to a cycle. Alvisi et al.~\cite{alvisi1999cicexp} have shown that a CIC protocol can handle cyclic communication patterns without the risk of a domino effect, but they may introduce significant overhead.

The most complete and well-documented CIC protocols are BCS~\cite{briatico1984BCS} and HMNR~\cite{hellary2000HMNR}. Initial tests indicate that the HMNR has better performance than BCS. Therefore, in this paper, we adopt HMNR as our CIC protocol. In short, in HMNR each operator keeps a Lamport clock and a vector clock plus three boolean vectors with a length equal to the number of operators participating in the pipeline. Every operator updates his Lamport clock by increasing its value when it takes a new checkpoint. The vector clock \textit{ckpt} stores how many checkpoints have been taken by each operator from the perspective of the current operator. A boolean vector \textit{sent\_to} keeps information about messages sent to other operators since the last checkpoint of the current operator. Another boolean vector \textit{taken} stores the existence of Z-paths since the last known checkpoint. The last boolean vector \textit{greater} stores the information whether the operator's clock is greater or not from each other operator's clock. The operator's Lamport clock, the vector clock \textit{ckpt}, the boolean vector \textit{taken}, and the boolean vector \textit{greater} are piggybacked to every message. The protocol uses all these structures to detect cycles and decide when to force a checkpoint. When an operator receives a message, it checks if there is a message previously sent from it to the sender and the sender's clock is larger than its own or if there is a Z-path detected in the current checkpoint interval of the sender operator. More details on the cycle detection and the forced checkpoints can be found in the original paper~\cite{hellary2000HMNR}.

\para{Strengths} The primary strength of the CIC protocol is the forced checkpoints mechanism, leading to a smaller rollback distance and, most importantly, eliminating the domino effect. 

\para{Drawbacks} The main drawback of a CIC protocol is the overhead it introduces. For big and complex pipelines, the vector clocks and the boolean vectors can be rather large and greatly impact the size of the messages flowing throughout the system. 

\vspace{-1mm}
\section{Testbed System}
\label{sec:system}

We compared the checkpointing protocols in Styx \cite{psarakis2024styx}, the backend of Stateflow \cite{stateflow}. For the requirements of our experiments, we developed all necessary protocol mechanisms (e.g., message logging and coordination) and streaming operators (i.e., map, filter, window, join, aggregates).

The Stateflow cluster consists of the typical architecture. A coordinator node is responsible for scheduling/deploying the dataflow graph to workers and running the coordination logic of the checkpointing protocols. Worker nodes execute the dataflow logic and take checkpoints asynchronously based on the checkpointing algorithm. Finally, Stateflow uses Apache Kafka as a replayable fault-tolerant source and Minio as a persistent state store for the operator state checkpoints.

We choose Stateflow for the following reasons: i) unlike other streaming dataflow systems such as ApacheFlink, Stateflow allows for cycles in the dataflow graph; ii) Stateflow provides a sandboxed environment, where we can evaluate the different protocols in isolation, without additional overhead; iii) Other systems (e.g., Apache Flink) base their entire design on coordinated checkpoints -- when implementing uncoordinated protocols on Apache Flink, we realized that we needed to virtually rewrite the complete system itself.

\section{Metrics}
\label{sec:metrics}

Although there is a significant body of work in benchmarking and evaluating stream processing systems and fault tolerance (\Cref{sec:related-work}), no metrics are established to measure the performance of a checkpointing protocol meaningfully. In this work, we argue that the following metrics should be used to evaluate the performance of such a protocol.

\para{End-to-end Latency} A standard metric to evaluate the performance of stream processing systems is the \emph{end-to-end latency}, i.e., the time it takes for a record to result into output in the sink from the moment it is available in the input queue. Although latency is mainly related to the deployed query and the underlying system rather than the checkpointing protocol itself, it allows us to measure the impact of each protocol on normal execution, as the overhead it introduces in terms of latency. We opt to measure the 50th and 99th percentiles.

\para{Sustainable Throughput} Another common metric in stream processing literature is the \emph{maximum sustainable throughput} \cite{benchmarkingICDE18}. The \emph{maximum sustainable throughput} indicates the maximum throughput that the system can handle for a long period of time without provoking backpressure. Backpressure leads to constantly increasing latencies and an average processing throughput that is lower than the rate of incoming messages. Similarly to end-to-end latency, it allows us to assess the impact of the checkpointing protocol on the overall performance.

\para{Average Checkpointing Time} In this work, we measure the \emph{average checkpointing time}, i.e., the average time it takes for each protocol to take a checkpoint. The fundamental differences between the protocols lie in checkpoint triggering and the additional information that needs to be captured apart from the internal state. Therefore, measuring how these differences affect the time it takes to capture a checkpoint is crucial. Also, as the checkpointing time rises, a significant impact on the processing performance is expected. 

\para{Restart \& Recovery Time} Restart time consists of all the time the system spends to reload all the needed states and be ready to process data. The recovery time, on the other hand, informs us how long it takes to recover from a failure. The measurement starts when the failure is detected and finishes when the system has managed to return to normal execution. The higher the recovery time, the bigger the impact of a failure. Recovery time also encompasses restart time. 

\para{Invalid Checkpoints} Depending on the checkpointing protocol, invalid checkpoints may exist, i.e., checkpoints that cannot be part of a consistent recovery line and, thus, cannot be used for recovering after failure. The existence of invalid checkpoints can be problematic as the state grows since a lot of expensive storage space is occupied by information that will never be used. Moreover, invalid checkpoints can lead to significant rollback distance, which will result in replaying and reprocessing a significant number of messages. Therefore, the number of invalid checkpoints is a good indicator of the performance of a checkpointing protocol. The fewer invalid checkpoints exist, the better a protocol is performing.  

\para{Message Overhead} Each protocol introduces messages and requires specific information to be exchanged between workers or sent to the coordinator. Measuring the size of protocol-related information that circulates the system during execution allows us to capture the overhead that the protocol introduces in network usage. A higher percentage of protocol-related information means that a significant portion of our network is used, and additional serialization/deserialization CPU time is spent on information unrelated to processing.

\section{Streaming Query Workload}
\label{sec:queries}

To evaluate the checkpointing protocols, we employ four distinct queries from NexMark~\cite{tucker2008nexmark} and our adaptation of the cyclic query introduced in \cite{chadramouli2009cyclic}. 

\para{NexMark Queries} NexMark benchmark~\cite{tucker2008nexmark} simulates an e-commerce application and provides streaming queries with different properties and needs. We selected the following four queries, which allow us to measure the performance and the impact of the checkpointing protocols in different conditions:

\begin{itemize}
    \item \emph{Query 1} is a stateless map query that transforms the bid values. There is no shuffling.
    \item \emph{Query 3} implements an incremental stateful join, which joins persons with auctions. It involves a complex topology and shuffling between operators.
    \item \emph{Query 8} employs a windowed join between users and auctions. We opt for a processing time tumbling window; however, the type of the time window does not affect the checkpointing protocol's performance. It employs a complex topology, shuffling, and the complexity of the windowing. To meaningfully measure the impact of the protocols on the latency during execution, we implement a running window, i.e., the processing is triggered on record arrival, and the window is cleaned when it expires.
    \item \emph{Query 12} employs a windowed count over bids. Similarly to query 8, we choose the running version of a processing time tumbling window. The query performs aggregation over time windows and includes minor shuffling.
\end{itemize}

Fundamental processing operators in modern stream processing engines \cite{hazelcast2021jet,carbone2015flink,akidau2015google} include maps, joins, windows, and aggregates. The queries we choose represent those fundamental operations and sufficiently cover the operations appearing in the NexMark suite.

\begin{figure}[t]
    \centering
    \includegraphics[width=0.8\linewidth]{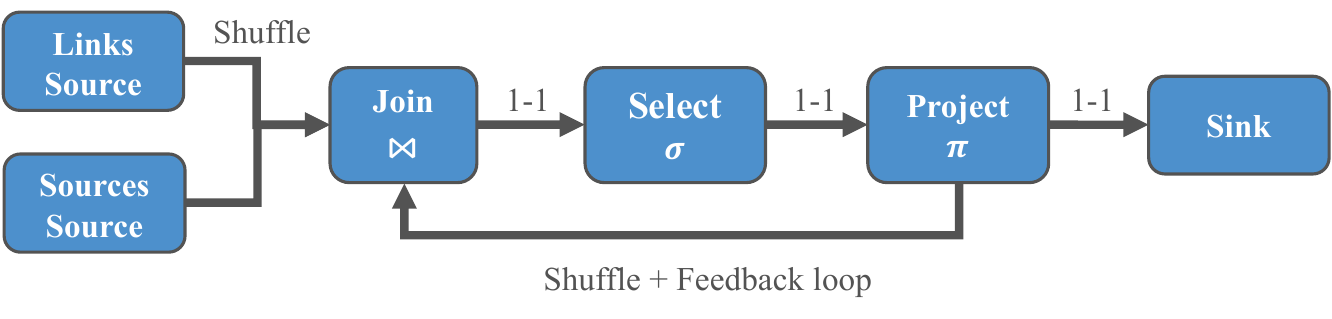}
    \caption{Execution graph of the reachability query.}
    \label{fig:reachability_query}
    \vspace{-3mm}
\end{figure}

\para{Cyclic Query} Most stream processing engines do not support cyclic queries. However, there is existing research on cyclic or recursive queries in stream processing~\cite{chadramouli2009cyclic, pacaci2020queryeval, mcsherry2013naiad}. To further enable research on cyclic streaming queries and to encourage stream processing engines to support such queries, it is essential to evaluate existing checkpointing protocols with cyclic queries. For our evaluation, we adapt the reachability query employed by FFP~\cite{chadramouli2009cyclic}. Given a static set of nodes, the goal of the query is to identify all reachable nodes from the available source nodes based on the available directed links between the nodes and provide the corresponding paths. The available source nodes and the directed links between the nodes are not known a priori, but they are processed on the fly and are temporal. \Cref{fig:reachability_query} illustrates the execution graph of the query. The query ingests two streams, the directed links between the nodes and the source nodes. Directed links are joined with sources that contain the starting node of the link as a reachable node. In the select operator, we check if the end node of the directed link of a joined pair is contained in the path of the source of the pair, and we discard such pairs. In the project operator, we discard unnecessary information and create a new source with the same source node, the end node of the link as a reachable node, and the path augmented by the pair's link. The new source is provided as output and recursively as input to the join operator. Finally, the join operator can receive direct messages when a specific link or source node is unavailable. In that case, it will remove every link or source affected from its state.

\section{Experimental Evaluation}
\label{sec:experiments}

\subsection{Evaluation setup}

The experiments are conducted on a local cluster with AMD EPYC 7H12 2.60GHz CPUs and 512GBs of memory. We deploy our benchmarking system using docker and docker-compose. Each worker uses 1 CPU for processing and handles a single parallel instance of each of the operators of the deployed pipeline. We do not use any limits on memory usage. Apache Kafka is used as the source and the sink of our system. Minio is used as a persistent storage for the checkpoints. We extend the NexMark generator from \cite{kalavri2018ds2, siachamis2023autoscalers} to provide the input in the required format of the system, and we provide a generator that creates source nodes and corresponding links for our cyclic query. We evaluate the three checkpointing protocols using the NexMark queries and our cyclic query. We implement and compare the vanilla versions of the protocols as described in \Cref{sec:checkpointing-protocols} in order to ensure a fair comparison of their core concepts that is not affected by optimizations tailored to specific system properties.

\subsection{Results}
In what follows, we present the results of our experimental evaluation of the three checkpointing protocols concerning the metrics for benchmarking checkpointing protocols that we previously discussed in \Cref{sec:metrics}. For the NexMark queries, we distinguish two settings: a balanced setting where the distribution of our input follows the uniform distribution and a skewed setting where we leverage NexMark's generator to provide different percentages of hot items.

\para{NexMark Queries} In practice, streaming systems are overprovisioned, ensuring a stable execution that does not cause backpressure in case of input rate fluctuations or transient system issues (e.g., garbage collection). In our experiments,  we run all queries at 80\% of the maximum sustainable throughput that each protocol achieves for each query and parallelism. We found 80\% to be the most stable configuration. Each run lasts for 60 seconds with 30 seconds of warmup. We introduce a failure on the 18th second of a 60-second run.

\begin{figure}[t]
    \centering
    \includegraphics[width=\linewidth]{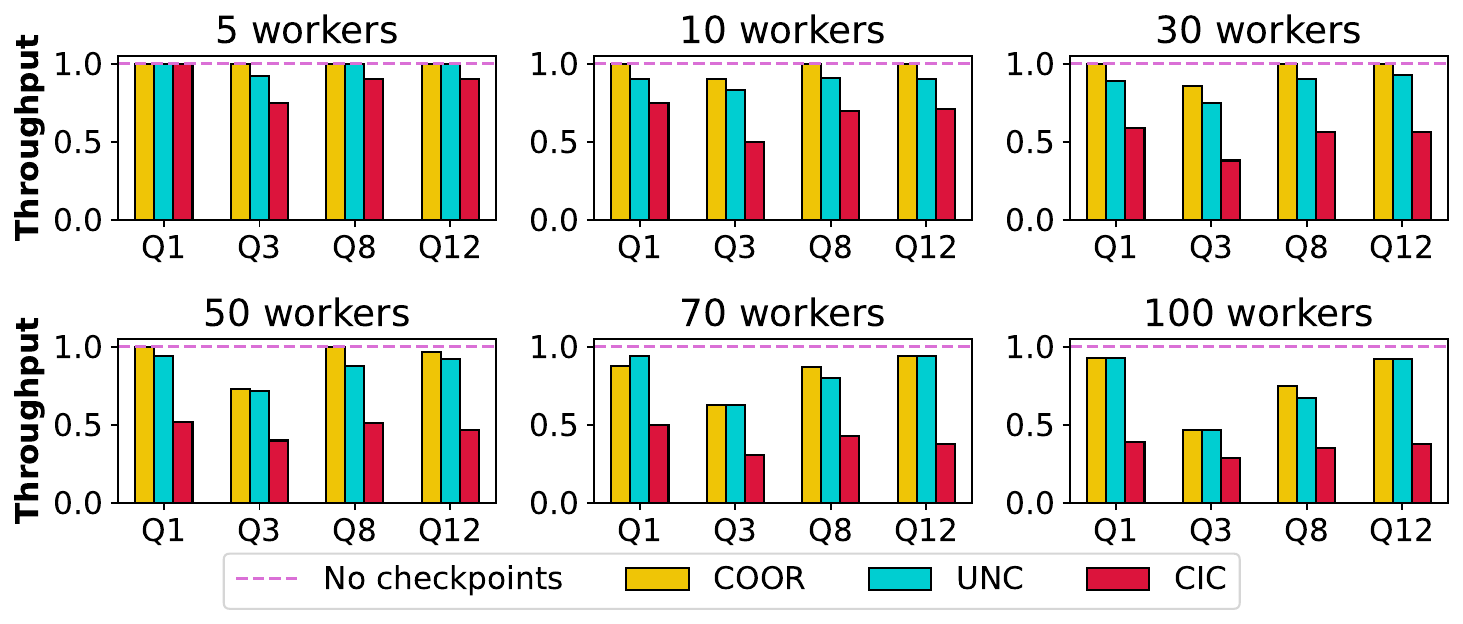}
    \caption{Normalized maximum sustainable throughput per query achieved by each protocol for different parallelism.}
    \label{fig:mst}
    \vspace{-3mm}
\end{figure}

\emphPara{Maximum Sustainable Throughput (MST)}
In \Cref{fig:mst}, we present the maximum sustainable throughput (MST) each protocol achieved normalized by the MST of the checkpoint-free execution for each query. For Q1, Q8, and Q12, the coordinated approach outperforms the rest and reaches the same MST as the checkpoint-free execution until we reach 70 workers. For 70 and 100 workers, we observe a slight decrease in MST for Q1 and Q12, which results in approximately 90\% of the checkpoint-free MST. The impact of the increase in parallelism is more significant for Q8, which employs a join. The uncoordinated protocol follows closely, achieving an MST around 10\% lower than the coordinated approach in all cases. On the other hand, the communication-induced protocol fails to keep up and, in higher parallelism, can reach an MST lower than 50\% of the checkpoint-free MST. None of the protocols can keep up with the checkpoint-free execution for Q3. However, the coordinated and uncoordinated protocols achieve an MST higher than 70\% of the optimal for Q3 in most cases while maintaining an MST of 50\% of the optimal for the edge case of 100 workers. On the contrary, the communication-induced protocol fails to achieve an MST higher than 50\% for Q3 primarily due to the high message overhead it introduces. In terms of MST, the coordinated approach outperforms the others, while only the uncoordinated can remain competitive.

\emphPara{Message Overhead}
The overhead of the protocol-related information transferred throughout the system can either be in the form of additional protocol messages and/or piggy-backed information to process messages. The only protocol-related overhead for the coordinated approach is the messages between workers and the coordinator when starting and concluding a coordinated round, and the markers forwarded from the sources to the pipeline sinks. The uncoordinated protocol requires the operators to send the metadata of every checkpoint they take to the coordinator. \Cref{tab:message-overhead} shows that the overhead that the coordinated and the uncoordinated introduce is insignificant in all cases.  On the contrary, as explained in \Cref{sec:checkpointing-protocols}, additionally to the information required by the uncoordinated protocol, the communication-induced protocol piggybacks to the process messages all the information required to decide on forcing a checkpoint. The size of this information depends on the number of total instances of the operators employed. As \Cref{tab:message-overhead} indicates, even for a parallelism of 10, the overhead can double the size of the messages that are communicated between the workers and the coordinator, while for 50 workers, the message size can reach up to 2.58x the size of messages of a checkpoint-free execution. Increased message size does not only result in the need for higher network bandwidth but also cripples the processing power of our system as it has to serialize and deserialize much larger messages. Therefore, it significantly affects the maximum sustainable throughput we can achieve using the communication-induced protocol, as shown in \Cref{fig:mst}.

\begin{figure}[t]
    \centering
    \includegraphics[width=\linewidth]{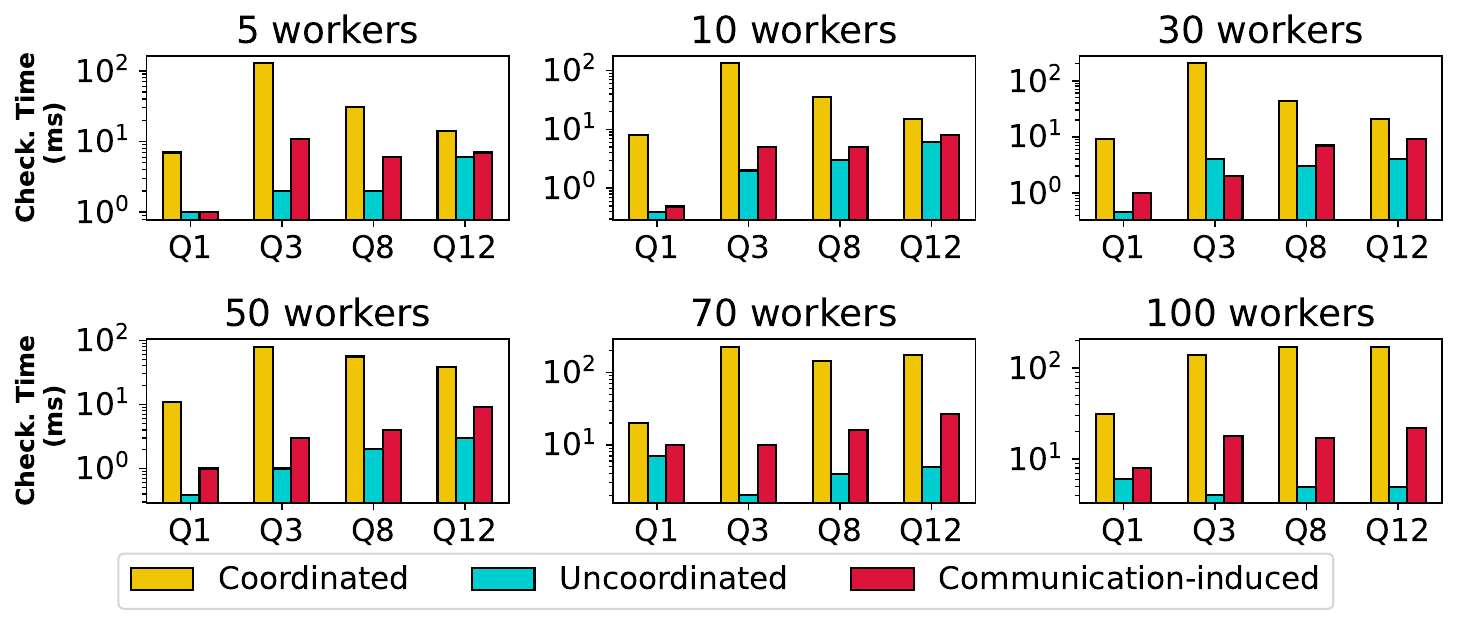}
    \caption{Average checkpointing time on different parallelisms.}
    \label{fig:avg_check_time}
    \vspace{-4mm}
\end{figure}

\begin{table}[t]
\centering
\caption{Ratio of message overhead with respect to an execution without checkpoints.\label{tab:message-overhead}}
\resizebox{\columnwidth}{!}{
\begin{tabular}{c|cccc|cccc|}
\cline{2-9}
\multicolumn{1}{l|}{}                       & \multicolumn{4}{c|}{10 workers}                                                          & \multicolumn{4}{c|}{50 workers}                                                          \\ \hline
\multicolumn{1}{|c|}{Protocol}              & \multicolumn{1}{c|}{Q1}   & \multicolumn{1}{c|}{Q3}   & \multicolumn{1}{c|}{Q8}   & Q12  & \multicolumn{1}{c|}{Q1}   & \multicolumn{1}{c|}{Q3}   & \multicolumn{1}{c|}{Q8}   & Q12  \\ \hline
\multicolumn{1}{|c|}{COOR}           & \multicolumn{1}{c|}{1.00x} & \multicolumn{1}{c|}{1.00x} & \multicolumn{1}{c|}{1.00x} & 1.00x & \multicolumn{1}{c|}{1.00x} & \multicolumn{1}{c|}{1.00x} & \multicolumn{1}{c|}{1.00x} & 1.00x \\ \hline
\multicolumn{1}{|c|}{UNC}         & \multicolumn{1}{c|}{1.00x} & \multicolumn{1}{c|}{1.00x} & \multicolumn{1}{c|}{1.00x} & 1.00x & \multicolumn{1}{c|}{1.00x} & \multicolumn{1}{c|}{1.01x} & \multicolumn{1}{c|}{1.01x} & 1.00 \\ \hline
\multicolumn{1}{|c|}{CIC} & \multicolumn{1}{c|}{2.10x} & \multicolumn{1}{c|}{1.82x} & \multicolumn{1}{c|}{1.74x} & 1.79x & \multicolumn{1}{c|}{2.53x} & \multicolumn{1}{c|}{2.58x} & \multicolumn{1}{c|}{2.49x} & 2.58x \\ \hline
\end{tabular}
}
\vspace{-3mm}
\end{table}

\begin{figure*}
    \centering
    \includegraphics[width=\textwidth]{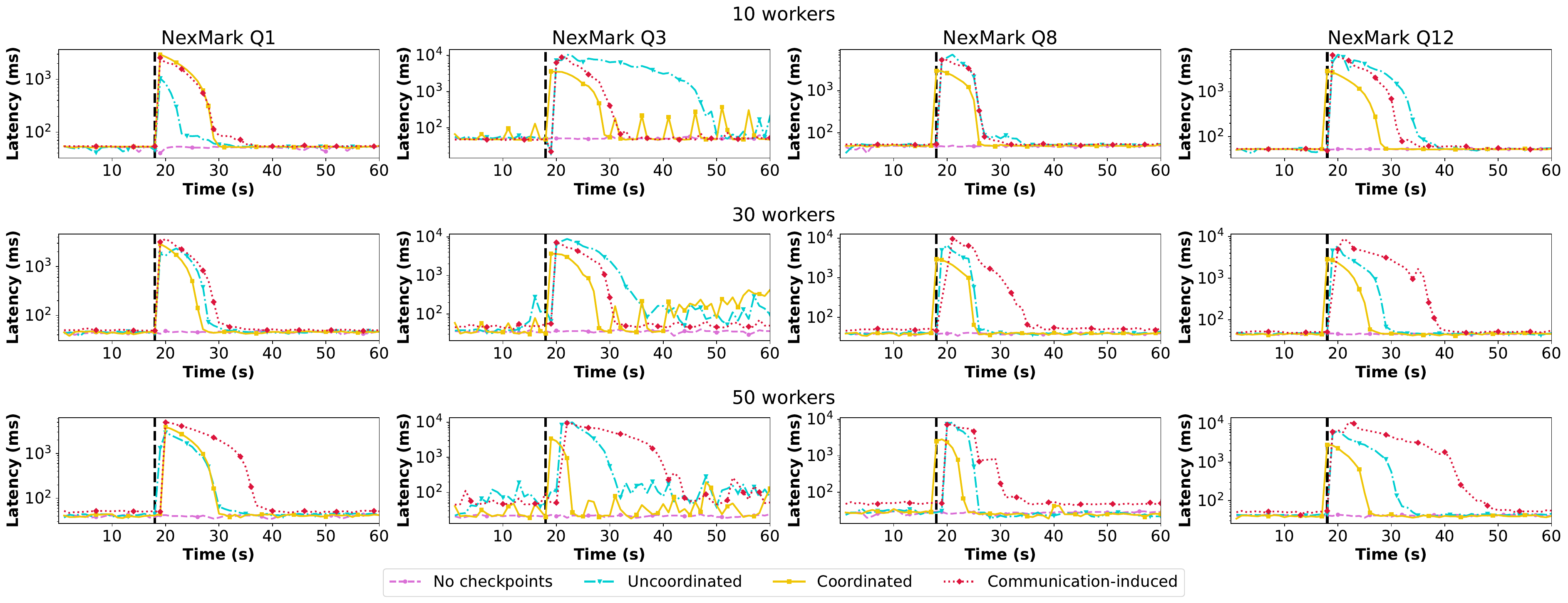}
    \caption{50th percentile latency. The black dashed vertical line indicates the moment of failure.}
    \label{fig:50th-latency}
    \vspace{-2mm}
\end{figure*}

\emphPara{Average Checkpointing Time}
We showcase the average checkpointing time for each protocol for all settings in \Cref{fig:avg_check_time}.
The uncoordinated and communication-induced protocols have an average checkpointing time of a few milliseconds for all settings. The coordinated approach requires a full checkpointing round to be completed to consider its checkpoints as valid. Therefore, in contrast to the other protocols, it incurs an average checkpointing time of up to two magnitudes higher for Q3, Q8, and Q12, which involve shuffling. This is especially the case for Q3, which employs a complex topology and has a high computational complexity, as well as for the higher parallelisms that result in a higher degree of shuffling. The latency overhead caused by the increased checkpointing time in Q3 is also visible in \Cref{fig:50th-latency} for 10 workers.

\begin{figure*}
    \centering
    \includegraphics[width=\textwidth]{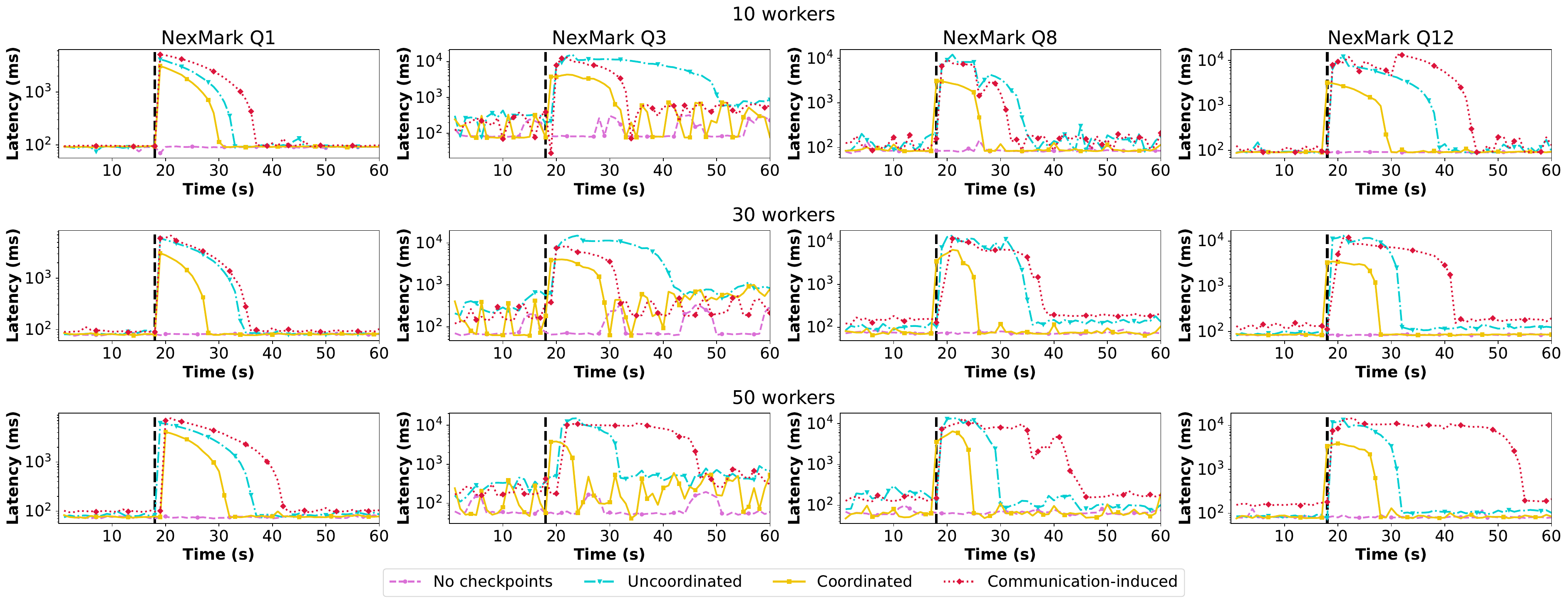}
    \caption{99th percentile latency. The black dashed vertical line indicates the moment of failure.}
    \label{fig:99th-latency}
    \vspace{-2mm}
\end{figure*}

\emphPara{Impact on the 50th and 99th percentile of latency}
In \Cref{fig:50th-latency} and \Cref{fig:99th-latency}, we present the 50th and 99th percentiles per second for each protocol and query for different parallelisms. Due to space limitations, we include 10, 30, and 50 workers in our discussion. However, the other settings follow a similar trend. The 50th percentile latency allows us to evaluate the mean performance of the protocols, while the 99th percentile highlights the stragglers and the outliers. For the settings of 10 and 30 workers, the 50th percentile for all protocols for the simpler queries Q1, Q8, and Q12 is similar before the failure occurred and after the system recovered to a stable execution. However, for the 50-worker case, the communication-induced protocol requires piggybacking additional protocol information of significant size at every message. This results in a slight increase observed in the 50th percentile, which is considerably higher in Q8 because it employs a costly join. As for Q3, the coordinated approach suffers from latency spikes every time a checkpoint is taken, which is more evident as the state grows and for the 10-worker case. The 99th percentile follows the same patterns as the 50th percentile for the execution period prior to the failure. Q3 employs an incremental join; the spikes and the increasing instability in latency that we observe are expected and attributed to a combination of the query's nature and checkpointing.

\emphPara{Recovery \& Restart Time}
Recovery time is the time passed from detecting the failure until the system returns to normal and stable execution.
Looking at the 50th percentile (\Cref{fig:50th-latency}), all protocols require around 10 seconds to recover to normal execution for Q1 for a parallelism of 10, while very small differences are also observed for Q1 for 30 workers. For 50 workers, the communication-induced protocol requires around 10 more seconds to recover due to the significant message overhead it introduces. For Q8 and Q12, the communication-induced protocol performs marginally better than the uncoordinated protocol for 10 workers, but it falls behind when the parallelism increases as it requires around 10 more additional seconds to recover. In Q3, the communication-induced protocol has a smaller recovery time than uncoordinated by up to 20 seconds for 10 and 30 workers, resulting from replaying fewer messages due to forced checkpoints closer to the failure. On the other hand, it requires 10 additional seconds for 50 workers. On average, the coordinated protocol greatly outperforms the other protocols regarding recovery time, mostly because the uncoordinated and communication-induced protocols have to replay many messages. 

\begin{figure}[t!]
    \centering
    \includegraphics[width=\linewidth]{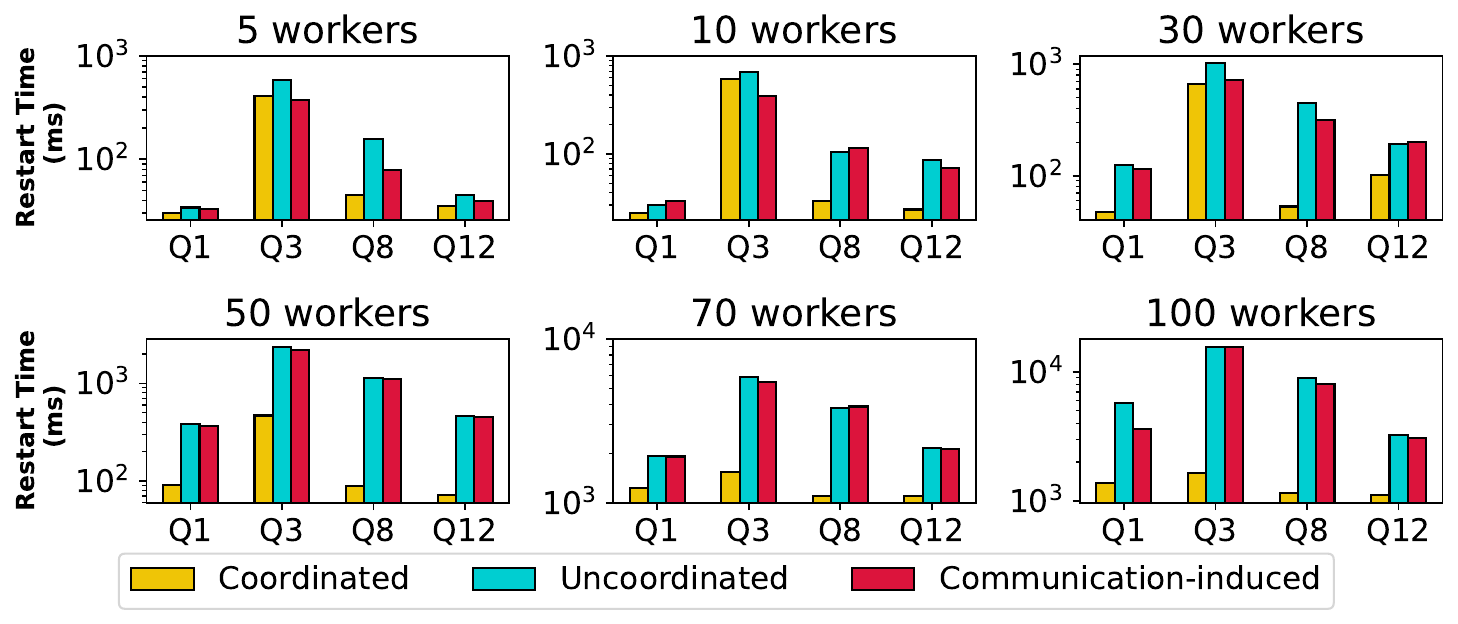}
    \caption{Restart time after failure per query for each protocol on different levels of parallelism.}
    \label{fig:50w-rt}
    \vspace{-4mm}
\end{figure}

The restart time (\Cref{fig:50w-rt}) is part of the recovery time and reflects the time passed from detecting the failure until the system is ready to restart processing. On average, the coordinated protocol restarts faster than the other two protocols. This is especially evident for a larger number of workers. For example, the restart process for the uncoordinated and communication-induced protocols can take up to 10 times longer than the coordinated for 100 workers. The UNC and CIC protocols need to fetch and prepare the messages to replay and, therefore, take more time to restart. On the other hand, finding the recovery line has an insignificant cost.

\begin{table}[t]
\centering
\caption{Total checkpoints and percentage of invalid checkpoints.\label{tab:num_checkpoints}\vspace{-3mm}}
\resizebox{\columnwidth}{!}{
\begin{tabular}{ccc|ccc|cccccc|}
\multicolumn{1}{l}{}         & \multicolumn{3}{c}{10 workers}  & \multicolumn{3}{c}{50 workers}  \\ \cline{2-7}                                                              
\multicolumn{1}{l|}{}         & \multicolumn{3}{c||}{\textit{\textbf{Total(Invalid)}}}   & \multicolumn{3}{c|}{\textit{\textbf{Total(Invalid)}}}  \\ \hline
\multicolumn{1}{|c||}{Query} & \multicolumn{1}{c|}{\textit{UNC}} & \multicolumn{1}{c|}{\textit{CIC}} & \multicolumn{1}{c||}{\textit{COOR}}  & \multicolumn{1}{c|}{\textit{UNC}}  & \multicolumn{1}{c|}{\textit{CIC}} & \multicolumn{1}{c|}{\textit{COOR}} \\ \hline \hline
\multicolumn{1}{|c||}{Q1}    & \multicolumn{1}{c|}{303(0\%)}  & \multicolumn{1}{c|}{285(0\%)} & \multicolumn{1}{c||}{240(0\%)}  & \multicolumn{1}{c|}{1437(0\%)} & \multicolumn{1}{c|}{1428(0\%)} & \multicolumn{1}{c|}{1200(0\%)} \\ \hline
\multicolumn{1}{|c||}{Q3}    & \multicolumn{1}{c|}{455(4\%)}  & \multicolumn{1}{c|}{471(3\%)} & \multicolumn{1}{c||}{400(0\%)}  & \multicolumn{1}{c|}{2399(3\%)} & \multicolumn{1}{c|}{2517(4\%)} & \multicolumn{1}{c|}{2000(0\%)} \\ \hline
\multicolumn{1}{|c||}{Q8}    & \multicolumn{1}{c|}{384(2\%)}  & \multicolumn{1}{c|}{386(3\%)} & \multicolumn{1}{c||}{360(0\%)}  & \multicolumn{1}{c|}{1924(2\%)} & \multicolumn{1}{c|}{1920(3\%)} & \multicolumn{1}{c|}{1800(0\%)} \\ \hline
\multicolumn{1}{|c||}{Q12}   & \multicolumn{1}{c|}{282(3\%)}  & \multicolumn{1}{c|}{282(4\%)} & \multicolumn{1}{c||}{240(0\%)}  & \multicolumn{1}{c|}{1446(3\%)} & \multicolumn{1}{c|}{1451(3\%)} & \multicolumn{1}{c|}{1200(0\%)} \\ \hline
\end{tabular}
}
\vspace{-4mm}
\end{table}

\emphPara{Invalid checkpoints}
The percentage of invalid checkpoints over the total checkpoints indicates how much the system rolled back. Low percentages show no domino effect and better utilization of the checkpointed state. The coordinated approach does not introduce any invalid checkpoints. \Cref{tab:num_checkpoints} shows that for all the acyclic queries, the uncoordinated and communication-induced protocols introduce very few invalid checkpoints and result in similar total checkpoints. Overall, the uncoordinated and communication-induced protocols result in more checkpoints than the coordinated protocol since every operator independently decides when to take a checkpoint based on its worker's clock.

\para{Skewed NexMark} Operating under a skewed workload usually results in workers straggling to process the excessive load they are responsible for. Although operating under such conditions is not preferable, avoiding it is not always feasible. Therefore, it is important to investigate how the different protocols perform under skew. To measure the impact of skew on the protocols' performance, we employ Q3, Q8, and Q12 under different hot item ratios provided by the NexMark generator. Q1 is not affected by skew as it involves non-keyed operations. Therefore, we omit it. We run Q3, Q8, and Q12 on 10 workers at 50\% and 80\% of the maximum sustainable throughput of the non-skewed execution of every protocol without introducing any failure. Both throughputs result in straggling workers. However, the latter stresses significantly more the system, resulting in fewer checkpoints taken and higher sensitivity to skew. We consider these settings representative of a real-world deployment, where overprovisioning is employed to handle spikes and unexpected skews. We employed three different hot item ratios to increase the skew gradually, from 10\% to 30\%. The straggling workers heavily affect the 99th percentile of latency, so we focus on the 50th percentile. We also report the average checkpointing time, as it is also heavily affected by the skew and can significantly affect the latency.

\begin{figure}[t]
    \begin{subfigure}{\linewidth}
        \centering
        \includegraphics[width=\linewidth]{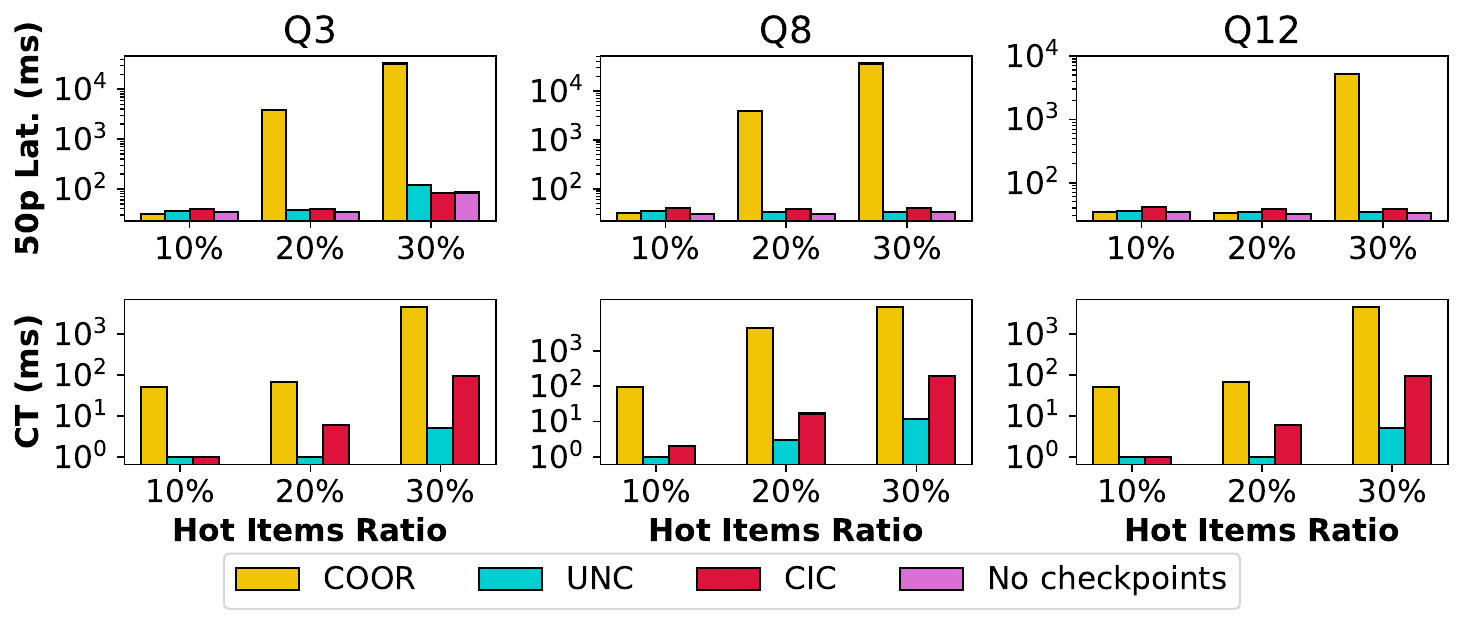}
        \caption{50\% of the MST of the non-skewed execution.}
        \label{fig:skew-50mst-all}
    \end{subfigure}
    \begin{subfigure}{\linewidth}
        \centering
        \includegraphics[width=\linewidth]{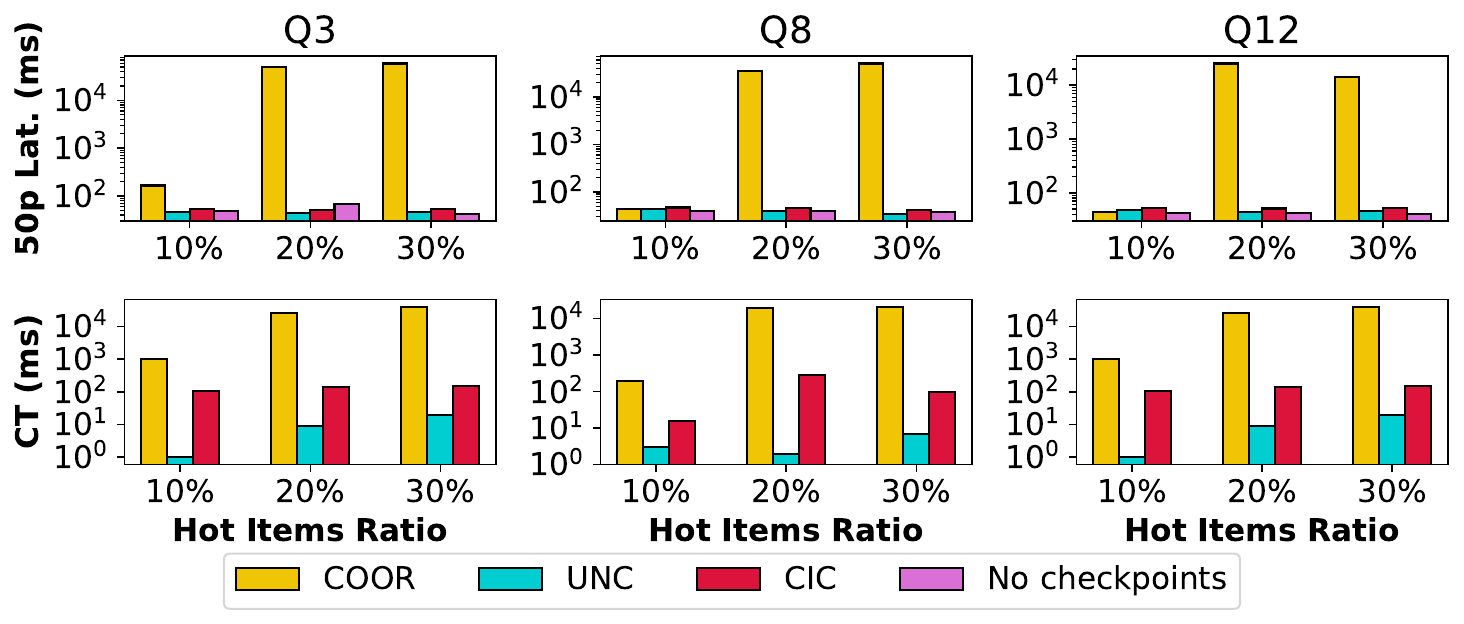}
        \caption{80\% of the MST of the non-skewed execution.}
        \label{fig:skew-80mst-all}
    \end{subfigure}
    \caption{50th percentile latency \& average checkpointing time under different hot items percentages.}
    \label{fig:skew-all}
    \vspace{-3mm}
\end{figure}

Unlike the non-skewed experiments, as illustrated in \Cref{fig:skew-all}, the coordinated protocol performs the worst regarding 50th percentile latency and average checkpointing time in both throughputs. With every increase in the hot items ratio, latency and checkpointing time increase by at least an order of magnitude for the lower throughput, while for the higher throughput, even the lowest skew ratio has a significant impact on Q3. The coordinated protocol is so heavily impacted by skew because not only are the straggling operators slow to take their checkpoints, but they also delay propagating their markers to downstream operators that block processing in other channels to wait for the delayed markers. Meanwhile, both UNC and CIC keep both metrics relatively low. In summary, the uncoordinated and communication-induced protocols can handle skew more effectively in every case.

\begin{figure}
    \centering
    \includegraphics[width=\linewidth]{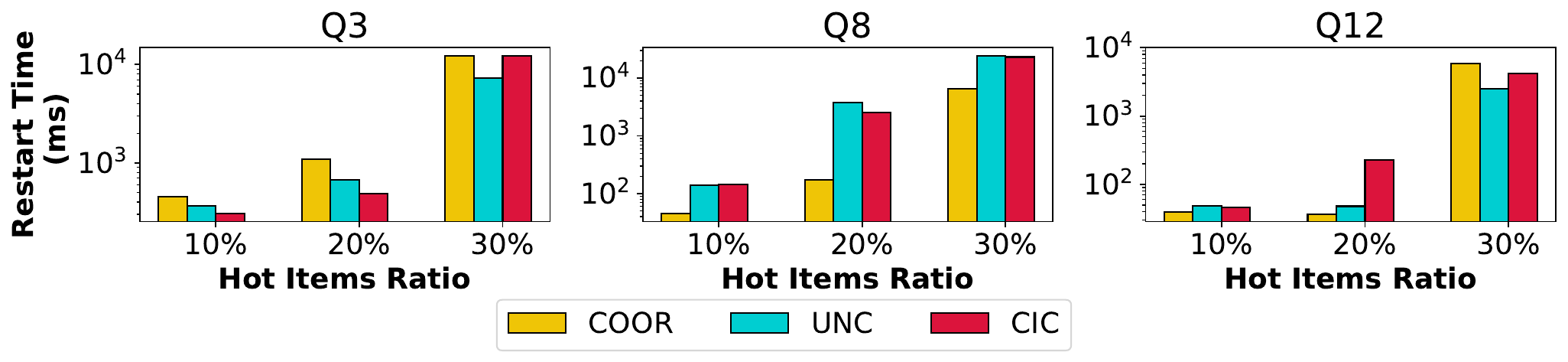}
    \caption{Restart time after failure per query in the presence of skew.}
    \label{fig:restart-skew}
    \vspace{-4mm}
\end{figure}

Similar to the non-skewed experiments, we perform another run using the 50\% MST, introducing a failure. \Cref{fig:restart-skew} shows the time needed to restart processing. Unlike the non-skewed experiments, where the coordinated outperformed the other approaches, the differences are mitigated under skew, and all protocols perform similarly. This is an immediate result of the coordination under skew with the stragglers. Invalid checkpoints remain the same under skewed and non-skewed conditions. We do not report recovery time since none of the protocols managed to recover within the time frame for 20\% and 30\% skew, while for 10\% skew, the performance is similar to the non-skewed experiments.

\para{Cyclic query} For the cyclic query, we only evaluate the \emph{uncoordinated} and the \emph{communication-induced} checkpointing protocols. The aligned version of the coordinated protocol cannot handle cyclic queries. That is because at least one operator would be waiting for a marker that originates from itself, thus leading to a deadlock. 

We evaluate the protocols with two parallelisms, 5 and 10 workers. We refrain from using higher parallelisms since CIC is greatly affected by complex topologies and higher parallelism, as shown in \Cref{fig:mst}. For both deployments, we use the same configuration for our generator. It creates events with the following probabilities: 60\% chance of creating a new link, 15\% of creating a source node, 20\% chance of deleting an existing link, and 5\% of deleting an existing source node. The generator also assumes a static set of 1M nodes. We evaluate the two protocols with an input rate of 75\% - 80\% of their MST for the query. We run the experiments for 60 seconds and introduce a failure at the 48th second.

\begin{table}[t]
\caption{Average checkpointing time (CT), restart time (RT), and invalid checkpoints (IC) for the cyclic query.\label{tab:cyclic-results}}
\resizebox{\columnwidth}{!}{
\begin{tabular}{c|ccc|ccc|}
\cline{2-7}
\multicolumn{1}{l|}{}                     & \multicolumn{3}{c||}{\textbf{Uncoordinated}}                                                                      & \multicolumn{3}{c|}{\textbf{Communication-induced}}                                                                      \\ \hline
\multicolumn{1}{|c|}{\textbf{\#Workers}}      & \multicolumn{1}{c|}{\textbf{CT}} & \multicolumn{1}{c|}{\textbf{RT}} & \multicolumn{1}{c||}{\textbf{IC}} & \multicolumn{1}{c|}{\textbf{CT}} & \multicolumn{1}{c|}{\textbf{RT}} & \multicolumn{1}{c|}{\textbf{IC}} \\ \hline
\multicolumn{1}{|c|}{5} & \multicolumn{1}{c|}{0.01 ms}       & \multicolumn{1}{c|}{620 ms}         & \multicolumn{1}{c||}{1.4\%}                                & \multicolumn{1}{c|}{2.73 ms}      & \multicolumn{1}{c|}{347 ms}         & 1.7\%                                \\ \hline
\multicolumn{1}{|c|}{10} & \multicolumn{1}{c|}{1.38 ms}        & \multicolumn{1}{c|}{344 ms}         & \multicolumn{1}{c||}{1.4\%}                                & \multicolumn{1}{c|}{8.39 ms}       & \multicolumn{1}{c|}{399 ms}          & 1.6\%                                \\ \hline
\end{tabular}
}
\vspace{-5mm}
\end{table}

Regarding latency and maximum sustainable throughput, both protocols perform similarly to a checkpoint-free execution; therefore, we omit these metrics. We present the average checkpointing time, the recovery time, and the number of invalid checkpoints in \cref{tab:cyclic-results}. Regarding average checkpointing time, the uncoordinated protocol is faster than the communication-induced protocol since the communication-induced protocol requires checkpointing additional protocol-related information apart from an operator's state. However, the difference between the two measurements is practically insignificant. The communication-induced protocol required less time to restart after a failure for a parallelism of 5 workers, as it forced checkpoints that led to fewer messages being prepared to be replayed. For 10 parallel workers, the uncoordinated protocol restarts slightly faster than the communication-induced protocol, although the difference is insignificant. Based on the literature and the core characteristics of both protocols, the uncoordinated protocol was expected to introduce many invalid checkpoints and lead to a domino effect. Although this might still hold in some extreme cases, our experiments show that both protocols unexpectedly share very similar percentages of invalid checkpoints for both parallelisms. Neither protocol outperforms the other when employed on top of cyclic queries in any meaningful aspect, and the uncoordinated protocol does not introduce a domino effect.

\para{Summary} In our experiments, we explore three different cases: the NexMark queries with a uniformly distributed workload, the three more complex NexMark queries, i.e., Q3, Q8, and Q12 for a skewed input, and a cyclic query. In the first case, the coordinated approach outperforms the rest regarding latency, recovery time, and maximum sustainable throughput but has a significantly higher checkpointing time. Surprisingly, in contrast to the theoretical analysis, although parallelism and shuffling impact the checkpointing time of the coordinated protocol, they hardly affect the overall performance and only result in mild spikes in latency when a checkpoint is taken. Additionally, the uncoordinated protocol remains competitive in all queries and parallelisms. However, under skewed inputs, the uncoordinated greatly outperforms the coordinated one, which suffers both in terms of latency and checkpointing time. For the cyclic query, surprisingly, the uncoordinated does not showcase an increased number of invalid checkpoints (e.g., a domino effect) and performs slightly better than the communication-induced.

\section{Related Work}
\label{sec:related-work}

This section presents the related work regarding \emph{benchmarking} for stream processing systems and \emph{experimental evaluation} of fault tolerance in stream processing.

\para{Benchmarking for stream processing}
Linear Road~\cite{arasu2004linearroad} is one of the first benchmarks proposed for stream processing that simulates a traffic monitoring application and evaluates the benchmarked solution in terms of latency, throughput, and accuracy. CityBench~\cite{ali2015citybench} and RioTBench~\cite{shukla2017riotbench} are real-time analytics benchmarks that employ real-world Internet of Things (IoT) data and extend the evaluation using metrics such as memory and CPU utilization and completeness of query results. SparkBench~\cite{li2015sparkbench} is tailored to Apache Spark and targets CPU and memory utilization, network and disk I/O, job execution time, and throughput. NEXMark~\cite{tucker2008nexmark} is a widely adopted benchmark, also extended by Apache Beam~\cite{beamNexmark}, represents an e-commerce application, and provides streaming queries that cover all the fundamental processing workloads. 

\para{Experimental evaluation of fault tolerance}
StreamBench~\cite{lu2014streambench} employs seven workloads on Spark and Storm and performs an evaluation focusing on throughput and latency.
Qian et al.~\cite{qian2016icit} evaluate fault tolerance, including additionally Samza and Kafka. However, their evaluation lacks representative workloads as they only consider a simple workload that consumes input and performs no operations.

\section{Conclusions}
\label{sec:conclusions}
In this paper, we surveyed the three checkpoint protocol families for fault-tolerance in stream processing and discussed the theoretical advantages and drawbacks of each one of them. We developed an open-source testbed system that allows for isolated comparison of the approaches and performed a thorough experimental evaluation. While our experiments empirically confirmed the reasons behind the universal adoption of the coordinated approach, they also highlighted cases (e.g., skewed input) where the uncoordinated approach shows more robustness and better performance. Based on these results, we urge the research community to further research the uncoordinated approach since even a "vanilla" implementation of it was proven to perform well in uniformly distributed workloads, and it is the only viable solution for skewed workloads.

\section*{Acknowledgment}

This publication is part of project number 19708 of the Vidi research program, partly financed by the Dutch Research Council (NWO). It is also partially funded by ICAI AI for Fintech Research Lab. We would like to thank Gianni Wiemers for his contribution to the early stages of development.

\bibliographystyle{plain}
\bibliography{references}

\end{document}